\documentclass[runningheads]{svmult}

\usepackage{makeidx}   % allows index generation
\usepackage{graphicx}  % standard LaTeX graphics tool
\usepackage{subeqnar}  % subnumbers individual equations
\usepackage{multicol}  % used for the two-column index
\usepackage{physprbb}  % modified textarea for proceedings,
\makeindex             % used for the subject index
\begin{document}
\title*{Globular Clusters in Compact Groups}
\toctitle{Globular Clusters in Compact Groups}
% allows explicit linebreak for the table of content
%
%
\titlerunning{Globular Clusters in Compact Groups}
% allows abbreviation of title, if the full title is too long
% to fit in the running head
%
\author{Cristiano Da Rocha\inst{1}
\and Claudia Mendes de Oliveira\inst{1}
\and Michael Bolte\inst{2}
\and \\ Bodo L. Ziegler\inst{3}
\and Thomas H. Puzia\inst{4}}
\authorrunning{C. Da Rocha et al.}
% if there are more than two authors,
% please abbreviate author list for running head
%
%
\institute{Instituto de Astronomia, Geof\'{\i}sica e Ci\^encias Atmosf\'ericas
 (IAG), \\ Univ. de S\~ao Paulo, Rua do Mat\~ao, 1226, 05508-900, S\~ao Paulo, 
 SP, Brazil
\and UCO/Lick Observatory, Department of Astronomy and Astrophysics, \\
  University of California, Santa Cruz, California 95064, USA
\and Universit\"atssternwarte G\"ottingen, Geismarlandstr. 11, 37083
  G\"ottingen, Germany
\and Universit\"atssternwarte M\"unchen, Scheinerstr. 1, D-81679
  M\"unchen, Germany}

\maketitle              % typesets the title of the contribution

\begin{abstract}
We have studied globular cluster systems (GCSs) around elliptical galaxies
in Hickson compact groups using multi--band deep, high quality images
from Keck, VLT and CFHT. Analyzing the luminosity functions, specific
frequencies, color and spatial distributions, we could determine the
properties of the GCSs of those galaxies and trace their star formation
histories. We have found poor populations, concentrated toward the
center of the galaxies, with bimodal color distributions. The study
of GCSs around galaxies in small groups are a blank on the globular
cluster literature.
\end{abstract}

\section{Introduction}

Compact groups of galaxies are high density environments with small
velocity dispersions where interactions should play a major role on
the galaxies' evolution. A sample of 100 compact groups was selected by
Hickson (1982), the Hickson Compact Groups (HCGs). N-body simulations
predict a collapse time smaller a fraction of a Hubble time (Barnes,
1989). The fact that they still exist in the Universe, has generated
many discussions on its nature. The explanations presented range from
chance alignments of galaxies in poor groups to a delay on the group
collapse by special distributions of dark matter and kinematics.

In order to address the evolutionary history of galaxies in this kind of
environment, we continued the study of globular cluster systems (GCSs),
well known as good tracers of major events of star formation. Physical
properties as its specific frequency, luminosity, spatial and color
distributions, can give us important information about the formation
and evolution of their host galaxies (Ashman \& Zepf, 1998; Da Rocha et
al., 2002), since we expect them to show signatures of the interaction
processes.

\section{The Sample}

In this work we have studied early--type galaxies on three HCGs (HCG 22,
68 and 93) and the loose group Telescopium (Da Rocha, 2002; Da Rocha et
al., 2002).

{\bf HCG 22A} -- brightest galaxy of the triplet HCG 22, classified
as an E2 galaxy, at $33.1~Mpc$, observed with Keck/LRIS in the $B$
and $R$ bands.

{\bf HCG 68A/B} -- two first ranked galaxies at the quintet HCG 68,
classified as S0 and E2, at $23.8~Mpc$, observed with CFHT/MOS in the
$R$ band. Pair in probable interaction, it was not possible to separate
the individual GCSs, so we have analyzed it as a single GCS centered at
the pair geometric center.

{\bf HCG 93A} -- brightest galaxy of the quartet HCG 93, an E1 galaxy,
at $67.0~Mpc$, observed with CFHT/SIS in the $R$ band. Clear signs of
interaction with other group members like a central dust spiral and a
shell around it.

{\bf NGC 6868}, the brightest galaxy of the loose group Telescopium,
an E2 galaxy, at $26.8~Mpc$, which has probably suffered a recent merger
in its central part (Hansen et al., 1991). This galaxy will be used as
our control sample.

\section{Data Reduction}

The bright galaxies were modeled and subtracted with STSDAS tasks ELLIPSE
and BMODEL, and the detection and photometry was performed with SExtractor
(Bertin \& Arnouts, 1996). We have performed Monte--Carlo simulations
to evaluate the images detection limits, completeness fraction function,
photometric errors and star--galaxy separation index (see Da Rocha et al.,
2002 for a detailed description).

The GC candidates were selected considering: the detection in both bands,
for the cases where we have images in $B$ and $R$; classification as
point--source; magnitudes within the beginning of the luminosity function
($M_V \sim -11.0$) and the detection limit of the image and $(B-R)_0$
within 0.7 and 2.1.

\section{Analysis and Results}

\subsection{HCG 22A}

The magnitude limits applied to HCG 22A were $22.0 \le B \le 25.5$ and
$20.0 \le R \le 24.0$. Using the radial distribution of selected objects
we can notice a high concentration of objects close to the center,
where the GC candidates will be selected, and a flat outer region,
used to estimate the background level.

The globular clusters luminosity function (GCLF) was defined as a Gaussian
with turnover magnitude at $-7.33\pm0.04$ and $\sigma = 1.40\pm0.05$. To
HCG 22A this value are $B = 26.2\pm0.3$ and $R = 24.6\pm0.3$, which
are fainter than our detection limits. For this reason we have left the
turnover and the dispersion fixed.

\begin{figure}
\begin{center}
\includegraphics[width=.49\textwidth]{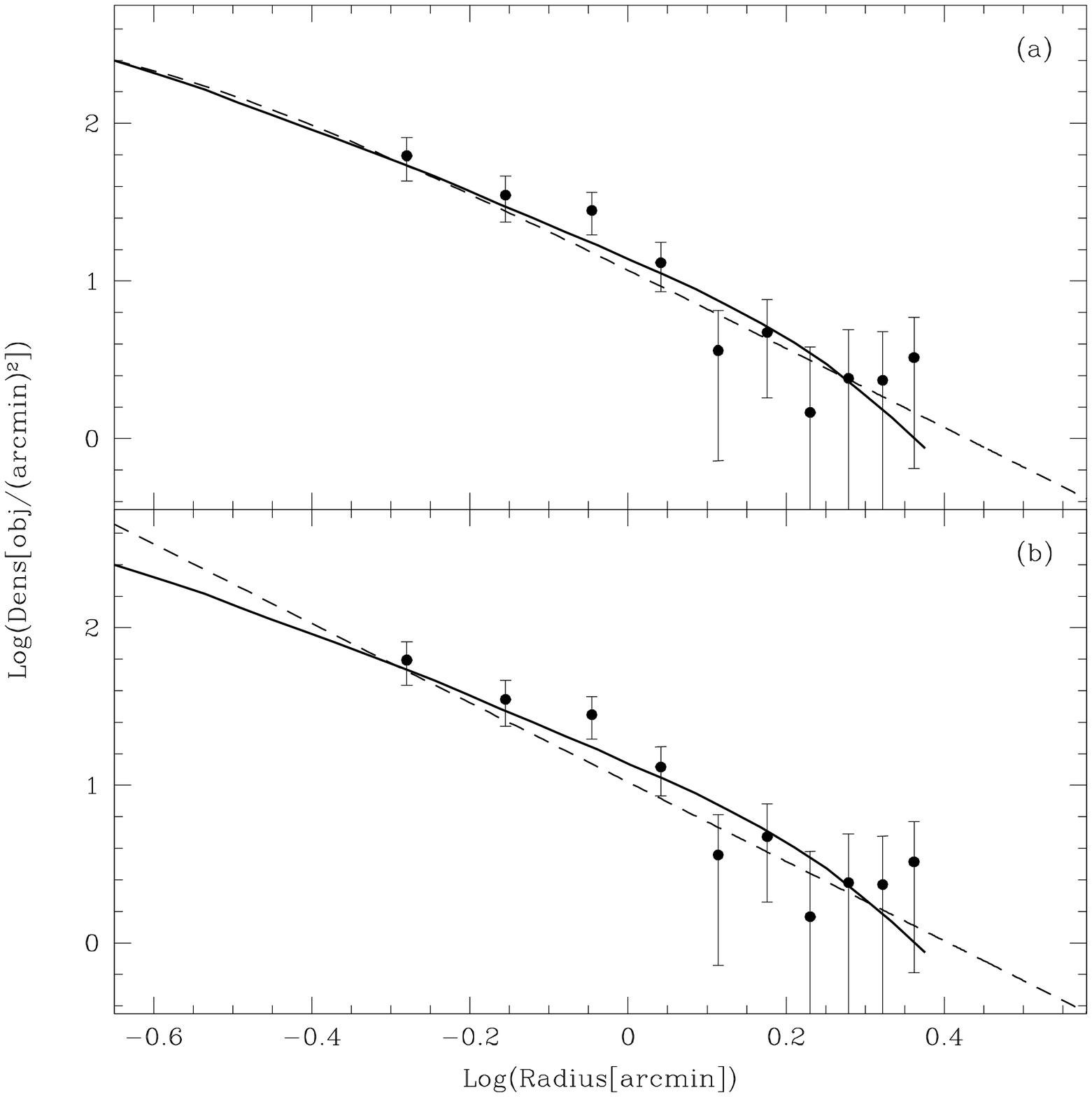}
\includegraphics[width=.49\textwidth]{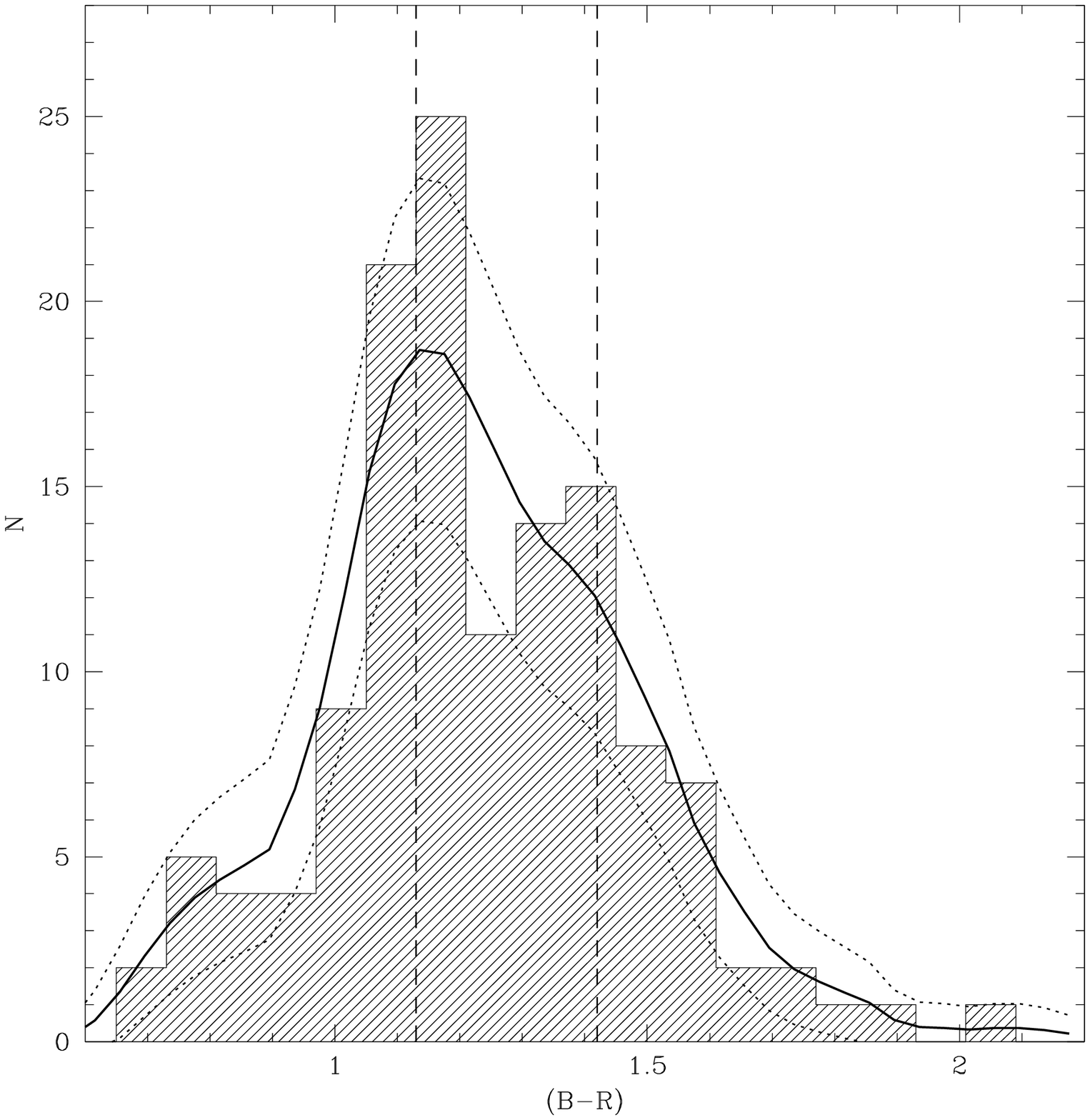}
\end{center}
\caption[]{The left panels show the radial distribution of GC candidates
around HCG 22A with the core model overplotted on dashed lines (panel
(a)), the power law model (panel (b)) and the galaxy light on continuous
line. The right panel shows the color distribution of the GC candidates
with Epanechnikov kernel density estimator and its upper and lower limits
overplotted (continuous and dotted lines). Dashed lines represent the
color peaks located by KMM.}
\label{figh22}
\end{figure}

To estimate the total number of objects in the GCS we have applied a
radial extrapolation using two radial models: a core model ($\rho \propto
(r_c^{\alpha} + r^{\alpha})^{-1}$) and a power law model ($\rho \propto
r^{-\alpha}$). We have obtained specific frequencies of $3.6\pm1.8$
and $5.2\pm3.2$ using the core and power law models, respectively,
showing a normal to poor GCS, considering the error bars. The normal $S_N$
value for an elliptical galaxy is $\sim 3.5$ (Harris, 2001). The high slope
fitted to the radial profile ($\alpha = 2.5\pm0.3$; see fig~\ref{figh22})
shows a GCS concentrated toward the center of the galaxy (the ``regular''
values range from $1.0$ to $2.0$ (Ashman \& Zepf, 1998)).

The color distribution of GCs around HCG 22A was analyzed with the
KMM (Ashman et al., 1994) which has detected the presence of two
sub--populations with peaks at $(B-R)_0 = 1.13\pm0.04$ and $1.42\pm0.04$
(see fig~\ref{figh22}), corresponding to metallicities of $[Fe/H] =
-1.45\pm0.21~dex$ and $-0.55\pm0.21~dex$. 62\% of the GCs belong to the
blue population and 38\% to the red one.

\subsection{NGC 6868}

To this ``control--sample'' galaxy the magnitude limits applied were
$22.0 \le B \le 24.5$ and $20.0 \le R \le 23.0$ and the GCLF turnover
is located at $B = 25.7\pm0.2$ and $R = 24.1\pm0.2$. The $S_N$ values
found for this galaxy were $1.8\pm1.1$ and $1.9\pm1.0$, for the core
and power law models, respectively, and show a poor GCS around this
galaxy. The radial profile has a regular value slope of $1.4\pm 0.3$.

The color distribution shows two sub--populations with peaks at $(B-R)_0
= 1.12\pm0.07$ and $1.42\pm0.07$ ($[Fe/H] = -1.48\pm0.22~dex$ and
$-0.55\pm0.22~dex$). 51\% of the GCs are part of the blue population
and 49\% of the red population.

\begin{figure}
\begin{center}
\includegraphics[width=.5\textwidth]{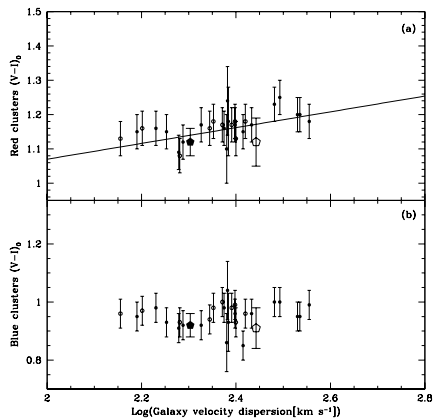}
\end{center}
\caption[]{Color--velocity dispersion relation for the early--type
galaxies of Forbes \& Forte (2001). Panel (a) shows the values for the
red peaks and panel (b) for the blue peaks. The solid line represents
the color--velocity dispersion relation proposed by those authors. The
filled pentagons are the values for HCG 22A and the open ones are the
values for NGC 6868.}
\label{figcomp}
\end{figure}

Comparing our results of HCG 22A and NGC 6868 with Forbes \& Forte (2001)
(fig~\ref{figcomp}), we can see that our results are consistent with
those authors'.

\subsection{HCG 68A/B}

The magnitude limits applied to this galaxy pair was $20.0 \le R \le
23.5$ and the GCLF turnover is located at $R = 23.8\pm0.3$, very close
to our detection limit. The high value estimated to the radial profile
($2.8\pm0.6$) shows, as in the case of HCG 22A, a GCS very concentrated
toward the geometric center of the galaxy pair. The $S_N$ values are
$1.3\pm0.8$ for the two models applied showing a poor GCS. Even if we
consider an upper limit case where the whole GCS would belong to the
elliptical galaxy, the specific frequency is not superior to $3.1\pm1.7$.

\begin{figure}
\begin{center}
\includegraphics[width=.5\textwidth]{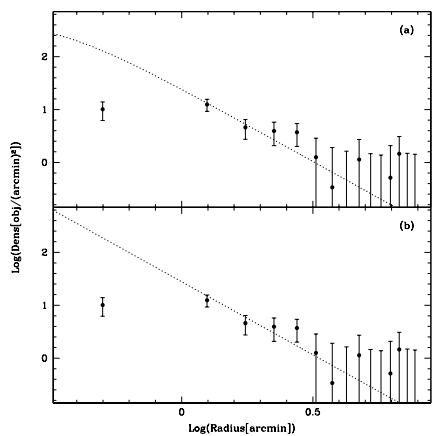}
\end{center}
\caption[]{Radial distribution of GC candidates around HCG 68A/B with
the core model overplotted (panel (a)) and power law model (panel (b)).}
\label{figh68}
\end{figure}

The flattening of the central region, that can be seen in the radial
profile (fig~\ref{figh68}), can be caused by the projection or by the
merger of the GCSs of the two galaxies and turns our $S_N$ estimated on
an upper limit.

\subsection{HCG 93A}

\begin{figure}
\begin{center}
\includegraphics[width=.5\textwidth]{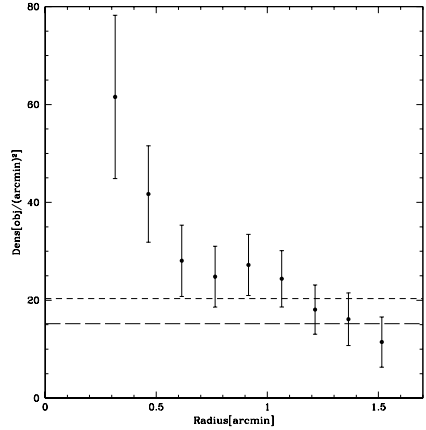}
\end{center}
\caption[]{Radial profile of selected objects around HCG 93A. Long dashed
line corresponds to the background level estimated with the flat part
from $0.^{\prime}69$ until $1.^{\prime}59$ (Analysis I). Short dashed
line corresponds to the background level estimated with the flat part
from $1.^{\prime}14$ until $1.^{\prime}59$ (Analysis II).}
\label{figh93}
\end{figure}

To this galaxy we have used $22.5 \le R \le 25.5$ as magnitude limits and
the GCLF turnover is located at $R = 26.3\pm0.3$. The radial profile of
selected objects shows a concentration of objects close to the center
and two different flat levels that could be used as background level
estimators (see fig~\ref{figh93}.

We have performed three cases of analysis: (1) the background level was
estimated using the whole flat area, from $0.^{\prime}69$ ($13~kpc$)
until $1.^{\prime}59$ ($31~kpc$); (2) we have estimated the background
level using the outer flat region, from $1.^{\prime}14$ ($22~kpc$) until
$1.^{\prime}59$; (3) no background estimative was used, all selected
objects were considered as GCs. The $S_N$ values estimated for the three
cases are extremely low (see table~\ref{tabh93}), showing a very poor
GCS. The slopes of each case can also be seen in table~\ref{tabh93},
and indicate a normal to high concentration GCS.

\begin{table}
\caption{Specific frequency values for HCG 93A}
\begin{center}
\renewcommand{\arraystretch}{1.4}
\setlength\tabcolsep{5pt}
\begin{tabular}{llll}
\hline\noalign{\smallskip}
Model & Analysis I & Analysis II & Analysis III\\
\noalign{\smallskip}
\hline
\noalign{\smallskip}
$S_N$ for core model & $0.16\pm0.11$ & $0.50\pm0.11$ & $1.57\pm0.11$ \\
$S_N$ for power law & $0.25\pm0.11$ & $0.50\pm0.11$ & $1.43\pm0.11$ \\
\noalign{\smallskip} 
\hline
Slope ($\alpha$) & $2.4\pm0.5$ & $1.4\pm0.2$ & $0.9\pm0.1$ \\
\hline
\end{tabular}
\end{center}
\label{tabh93}
\end{table}

\section{Summary}

In our study we have found poor GCSs in groups, compact or loose, showing
low specific frequency values. The GCS of the HCG galaxies are in all
cases very concentrated toward the center of the galaxy ($\alpha > 2.0$)
and the two multi--band analysis have shown multiple populations of GCs
around those galaxies.

A possible cause for those poor systems is the erosion caused by
the frequent interactions suffered by the galaxies in such dense
environments, which, associated to the expansion suffered by the GCS
during the interactions, may create a very efficient effect. Other
possible cause would be an inefficiency on forming GCs, comparing to
field stars, which may be due to the destruction of the GCs formation
loci, as the super giant molecular clouds.

We intent to enhance our studied sample in order to obtain a statistically
representative sample, enriching the literature, since there are only
a few studies of GCSs on the small groups environment, compact or loose.

\section{Acknowledgments}
C.D.R. and C.M.dO acknowledge funding from FAPESP (grant No. 96/08986-5),
PRONEX and the Alexander von Humboldt Foundation which made possible
the attendance to the conference and also to ESO and the conference
organizers for all the support.

%INDEX%%%%%%%%%%%%%%%%%%%%%%%%%%%%%%%%%%%%%%%%%%%%%%%%%%%%%%%%%%%%%%%
% Please check with the editor of your book whether he plans to
% include a "mutual" subject index - if so, please code your entries
% in the standard syntax. For your own purposes you may print your
% "personal" index by using the following commands:
%
%\clearpage
%\addcontentsline{toc}{section}{Index}
%\flushbottom
%\printindex
%%%%%%%%%%%%%%%%%%%%%%%%%%%%%%%%%%%%%%%%%%%%%%%%%%%%%%%%%%%%%%%%%%%%%

\end{document}